\documentclass[12pt,preprint]{aastex}

\shortauthors{Parrish et al.}

\begin{document}

\title{Simulation of the Magnetothermal Instability}

\author{Ian J. Parrish and James M. Stone\altaffilmark{1}}
\affil{Department of Astrophysical Sciences, Princeton University, Princeton, NJ 08544}

\altaffiltext{1}{Program in Applied and Computational Mathematics, Princeton University, Princeton, NJ, 08544}

\begin{abstract}
In many magnetized, dilute astrophysical plasmas, thermal conduction occurs almost exclusively parallel to  magnetic field lines.  In this case, the usual stability criterion for convective stability, the Schwarzschild criterion, which depends on entropy gradients, is modified.  In the magnetized long mean free path regime, instability occurs for small wavenumbers when $(\partial P/\partial
z)(\partial \ln T / \partial z) > 0$, which we refer to as the Balbus criterion.  We refer to the convective-type instability that results as the magnetothermal instability (MTI).  We use the equations of MHD with anisotropic electron heat conduction to numerically simulate the linear growth and nonlinear saturation of the MTI in plane-parallel atmospheres that are unstable according to the Balbus criterion.  The linear growth rates measured from the simulations are in excellent agreement with the weak field dispersion relation.  The addition of isotropic conduction, e.g. radiation, or strong magnetic fields can damp the growth of the MTI and affect the nonlinear regime.  
The instability saturates when the atmosphere becomes isothermal as the source of free energy is exhausted.  By maintaining a fixed temperature difference between the top and bottom boundaries of the simulation domain, sustained convective turbulence can be driven.  MTI-stable layers introduced by isotropic conduction are used to prevent the formation of unresolved, thermal boundary layers.  We find that the largest component of the time-averaged heat flux is due to advective motions as opposed to the actual thermal conduction itself. Finally, we explore the implications of this instability for a variety of astrophysical systems, such as neutron stars, the hot intracluster medium of galaxy clusters, and the structure of radiatively inefficient accretion flows.
\end{abstract}
\keywords{accretion, accretion disks --- convection --- hydrodynamics --- instabilities --- MHD --- stars: neutron --- turbulence}

\section{Introduction} \label{introduction}
In many dilute, magnetized astrophysical plasmas, the electron mean free path between collisions can be many orders of magnitude larger than the ion gyroradius.  In this regime, the equations of ideal magnetohydrodynamics (MHD) that describe the fluid plasma must be supplemented with anisotropic transport terms for energy and momentum due to the near free-streaming motions of particles along magnetic field lines \citep{brag65}.  Thermal conduction is dominated by the electrons compared to ions by a factor of the square root of the mass ratio.  In a non-rotating system, it is sufficient to neglect the ion viscosity \citep{bal04}.  A fully collisionless treatment may be done using more complex closures such as \citet{hp90}.  

The implications of anisotropic transport terms on the overall dynamics
of dilute astrophysical plasmas is only beginning to be explored
\citep{bal01, qdh02, sha03}.  One of the most remarkable results obtained thus
far is that the convective stability criterion for a weakly magnetized
dilute plasma in which anisotropic electron heat conduction occurs is
drastically modified from the usual Schwarzschild criteria \citep{bal00}.
In particular, stratified atmospheres are unstable if they contain a
{\em temperature} (as opposed to 
{\em entropy}) profile which is decreasing upward.  There are intriguing
analogies between the stability properties of rotationally supported
flows (where a weak magnetic field changes the stability criterion from a
gradient of specific entropy to a gradient of angular velocity), and the
convective stability of stratified atmospheres (where a weak magnetic
field changes the stability criterion from a gradient of entropy to a
gradient of temperature).  The former is a result of the magnetorotational
instability \citep[MRI;][]{bh98}.  The latter is a result of anisotropic
heat conduction.  To emphasize the analogy, we will refer to  this new
form of convective instability as the magnetothermal instability (MTI).
The MTI may have profound implications for the structure and dynamics
of many astrophysical systems.

In this paper, we use numerical methods to explore the nonlinear
evolution and saturation of the MTI in two-dimensions.  We adopt
an arbitrary vertical profile for a stratified atmosphere in which
the entropy increases upward (and therefore is stable according to
the Schwarzschild criterion), but in which the temperature is decreasing
upwards (and therefore is unstable according to the Balbus criterion, $(\partial P/\partial
z)(\partial \ln T / \partial z) < 0$).
We confirm the linear
growth rates predicted by \citet{bal00} for dynamically weak magnetic
fields and numerically measure the growth rates for stronger fields.  We find that in the nonlinear regime vigorous convective turbulence results in efficient heat transport.  Full details are published in \citet{ps05}, hereafter PS.  

These results may have implications for stratified atmospheres where anisotropic transport may be present.  The most exciting potential application is to the hot X--ray emitting gas in the intracluster medium of galaxy clusters \citep{pf05, mark98}.  The hot plasmas are magnetized and have very long mean free paths along the magnetic field lines; thus, they are a prime candidate for this instability. For example, the Hydra A cluster has $T \approx 4.5$ keV and a density of $n \approx 10^{-3} - 10^{-4} \mathrm{cm}^{-3}$ giving a mean free path that's almost one-tenth of the virial radius. Other applications are to the atmospheres of neutron stars with moderate magnetic fields and radiatively inefficient accretion flows.

\section{Physics of the MTI}
\subsection{Equations of MHD and Linear Stability}
The physics of the MTI is described by the usual equations of ideal MHD with the addition of a heat flux, $\mathbf{Q}$, and a vertical gravitational acceleration, $\mathbf{g}$.  
\begin{equation}
\frac{\partial \rho}{\partial t} + \mathbf{\nabla}\cdot\left(\rho \mathrm{\mathbf{v}}\right) = 0,
\label{eq:MHD_continuity}
\end{equation}
\begin{equation}
\frac{\partial(\rho\mathbf{v})}{\partial t} + \mathbf{\nabla}\cdot\left[\rho\mathbf{vv}+\left(p+\frac{B^{2}}{8\pi}\right)\mathbf{I} -\frac{\mathbf{BB}}{4\pi}\right] + \rho\mathbf{g}=0,
\label{eq:MHD_momentum}
\end{equation}
\begin{equation}
\frac{\partial\mathbf{B}}{\partial t} + \mathbf{\nabla}\times\left(\mathbf{v}\times\mathbf{B}\right)=0,
\label{eq:MHD_induction}
\end{equation}
\begin{equation}
\frac{\partial E}{\partial t} + \mathbf{\nabla}\cdot\left[\mathbf{v}\left(E+p+\frac{B^{2}}{8\pi}\right) - \frac{\mathbf{B}\left(\mathbf{B}\cdot\mathbf{v}\right)}{4\pi}\right] 
+\mathbf{\nabla}\cdot\mathbf{Q} +\rho\mathbf{g}\cdot\mathbf{v}=0,
\label{eq:MHD_energy}
\end{equation}
where the symbols have their usual meaning with $E$ the total energy.  The heat flux contains contributions from electron motions (which are constrained to move primarily along field lines) and isotropic transport, typically due to radiative processes.    Thus, 
${\bf Q} = {\bf Q}_{C} + {\bf Q}_{R}$, where 
\begin{equation}
\mathbf{Q}_{C} = - \chi_{C} \mathbf{\hat{b}\hat{b}}\cdot\mathbf{\nabla}T,
\label{eq:coulombic}
\end{equation}
\begin{equation}
\mathbf{Q}_{R} = - \chi_{R} \mathbf{\nabla}T,
\label{eq:radiative}
\end{equation}
where $ \chi_{C} $ is the Spitzer Coulombic conductivity \citep{spitz62},
$\mathbf{\hat{b}}$ is a unit vector in the direction of the magnetic field,
and $ \chi_{R} $ is the coefficient of isotropic conductivity.

Some progress can be made analytically in the linear regime of the instability.  I introduce two useful quantities,
\begin{equation}
\chi_{C}' = \frac{\gamma - 1}{P}\chi_{C} \qquad \textrm{and} \qquad \chi' = \frac{\gamma - 1}{P}\left(\chi_{C}+\chi_{R}\right).
\label{eq:cond_normalization}
\end{equation}
With WKB theory, one is able to obtain a dispersion relation.  The details of this process are to be found in \S 4 of \citet{bal00}.  The most important result of the linear analysis is the instability criterion,
\begin{equation}
k^2v_{A}^{2} - \frac{\chi'_{C}}{\rho \chi'}\frac{\partial P}{\partial z}\frac{\partial \ln T}{\partial z} < 0,
\label{eq:MTI_instability_criterion1}
\end{equation}
where $v_{A}^2=B^2/4\pi\rho$ is the Alfv\'{e}n speed.  In the limit of infinitesimal wavenumber, the instability criterion shows that any atmosphere with the temperature and pressure gradients in the same direction is unconditionally unstable, \textit{i.e.}
\begin{equation}
\frac{\partial P}{\partial z}\frac{\partial \ln T}{\partial z} > 0.
\label{eq:MTI_instability_criterion2}
\end{equation}
We refer to the instability criterion Eqn.[\ref{eq:MTI_instability_criterion2}] as the Balbus criterion.  The instability criterion of the magnetorotational instability \citep{bh98} can be written
\begin{equation}
k^2v_{A}^2 + \frac{d\Omega}{d\ln R} > 0,
\label{eq:MRI_instability_criterion}
\end{equation}
where $\Omega$ is the angular velocity.  The similarity between the MRI and the MTI is self-evident.  Strong magnetic fields are capable of stabilizing short-wavelenth perturbations in both instabilities through magnetic tension.
\subsection{Computational Method}
We use the 3D MHD code ATHENA \citep{gs05} with the addition of an operator-split anisotropic thermal conduction module for our simulations. Our initial state is always a convectively stable state ($dS/dz > 0$) in hydrodynamic equilibrium.  We implement two different boundary conditions for exploring the nonlinear regime.  The first boundary condition is that of an adiabatic boundary condition at the upper and lower boundaries, i.e. a Neumann boundary condition on temperature.  This situation is ideal for single-mode studies of linear growth rates.  In the Neumann boundary condition the magnetic field is reflected at the upper and lower boundaries, consistent with the adiabatic condition on heat flow. The second boundary condition fixes the the temperature at the upper and lower boundaries of the atmosphere, i.e. a Dirichlet boundary condition on temperature.  This setup is useful for driven simulations where we wish to study the effects of turbulence.  In this boundary condition, the magnetic field is again reflected at the boundaries of the box, but heat is permitted to flow across the boundary from the constant temperature ghost zones.  With these choices of boundary conditions, the net magnetic flux penetrating the box is constant in time as there is zero Maxwell stress at the boundary.  If the net magnetic flux penetrating the box is initially zero, then in two dimensions the magnitude of the magnetic field must decay in time as a result of Cowling's anti-dynamo theorem, but otherwise the saturated state is not affected. 
\section{Results}
\subsection{Single-Mode Perturbation and Qualitative Understanding of the MTI}
By examining a single mode-perturbation to the background state, we can gain an intuitive understanding of the physical mechanism of the instability.  We begin by perturbing a convectively-stable, but MTI-unstable atmosphere with a very weak sub-sonic and sub-Alf\'{e}nic sinusoidal velocity perturbation in a box with adiabatic boundary conditions. The evolution of the magnetic field lines is shown in Figure 1 for several different times.  Notice in the upper right plot that at $x\approx 0.03$ that a parcel of fluid (as traced by the frozen-in field line) has been displaced upward in the atmosphere.  As this parcel of fluid comes to mechanical equilibirum with the background state, it adiabatically cools.  The magnetic field line now is partially aligned with the background temperature state, thus thermally connecting this parcel of fluid with a hotter parcel deeper in the atmosphere.  As a result, heat flows along the field, causing the higher parcel to become buoyant.  This buoyant motion causes the field line to be more aligned with the background temperature, increasing the heat flux, and generating a runaway instability.  

It is instructive to examine the behavior of the temperature profile of the atmosphere.  Figure 2 shows the horizontally averaged temperature profile of this run at various times (normalized to the sound crossing time).  The initial state is a linear temperature profile decreasing with height.  As the instability progresses, the temperature profile becomes more and more isothermal.  Saturation occurs, as one would expect, when the temperature profile is almost completely isothermal, since the source of free energy has been depleted. More details are to be found in \S 4 of PS.
\subsection{Linear Growth Rates}
By following \S 4.3 of \citet{bal00} and using the Fourier convention, $\exp(\sigma t+ ikx)$, one can derive a weak-field dispersion relation for the MTI as
\begin{equation}
\left(\frac{\sigma}{N}\right)^{3} + \frac{1}{\gamma}\left(\frac{\sigma}{N}\right)^{2}\left(\frac{\chi' T k^2}{N}\right) +\left(\frac{\sigma}{N}\right) + \frac{d \ln T}{d \ln S}\left(\frac{\chi'_{c} T k^2}{N}\right) = 0,
\label{dispersionrelation}
\end{equation}
where $N$ is the Brunt-V\"ais\"al\"a frequency, the natural frequency of adiabatic oscillations for an atmosphere.  With the single-mode perturbation simulations we are able to measure the growth rate for a variety of situations.  Figure 3 plots the nondimensionalized growth rate versus wavenumber for theory (solid line) and the measured values from simulations (crosses).  As can be seen these are in very good agreement.

There are essentially two ways to suppress the growth of the MTI.  First, strong magnetic fields can exert tension that limits the growth and saturation of the instability.  Second, isotropic conduction, as would result from radiative transport, can effectively short-circuit the thermal driving along field lines necessary for this instability to occur.  For more detailed analysis, we refer the reader to \S 3 of PS.  
\subsection{Nonlinear Regime and Efficiency of Heat Transport}
In order to assess the efficiency of heat transport in the magnetothermal instability, one needs to examine multimode simulations seeded with Gaussian white noise perturbations and conducting boundary conditions at the top and bottom of the domain.  The simplest such set-up results in narrow, unresolved boundary layers at the upper and lower boundaries, thus, making the heat flux difficult to measure accurately.  As an alternative, we utilize a more physically relevant simulation.  This setup involves an atmosphere that is convectively stable throughout, but MTI unstable only in the central region.  The surrounding regions are stabilized to the MTI through the addition of isotropic conducitivity.  As a result, the central unstable region is well-resolved.  Figure 4 shows the evolution of magnetic field lines as this instability progresses.  The magnetic field lines shown essentially track the central unstable region; however, the third panel clearly shows a plume of fluid that is penetrating into the stable layer as a result of convective overshoot.  This phenomenon is well-known in the solar magnetoconvection literature \citep{tob01, bru02}.  In three dimensions this type of behavior greatly amplifies the magnetic field in a local magnetic dynamo.  

More quantitative measurements can be made by comparing the time- and horizontally-averaged vertical heat fluxes and breaking it down into Coulombic, radiative, and isotropic components.  Figure 5 shows these quantities plotted as a function of time.  At the midplane, the oscillatory advective flux is clearly dominant at any given instant in time; however, averaged in time the advective heat flux contributes roughly $\frac{2}{3}$ of the total heat flux.  The Coulombic flux, which is relatively constant in time, contributes the remaining $\frac{1}{3}$.  To determine the heat conduction efficiency of this instability, we compare it to the expected vertical heat flux across the simulation domain for pure uniform isotropic conductivity, namely, $Q_0 \approx 3.33\times 10^{-5}$.  The time-averaged heat conduction at the midplane is $\left<Q_{tot, 50\%}\right>\approx 3.54 \times 10^{-5}$, which indicates
that the instability transports the entire applied heat flux efficiently. 

\section{Conclusions and Application}
The most important conclusion of this work is that atmospheres with $dS/dZ < 0$ are not necessarily stable to convection.  In fact, dilute atmospheres with weak to moderate magnetic fields can be convectively unstable by the Balbus criterion resulting in an instability that we call the magnetothermal instability.  We have verified using MHD simulations that the measured linear growth rates agree with analytic WKB theory as predicted by \citet{bal00}.  For adiabatic boundary conditions, we find the saturated state is an isothermal temperature profile, corresponding to the exhaustion of the free energy in the system.  For a driven instability with conducting boundary conditions, we find that the MTI efficiently transports heat, primarily by advective motions of the plasma in the vigorous convection that results.

The most promising application of this instability is to clusters of galaxies.  Structure formation calculations assuming $\mathrm{\Lambda}$CDM cosmologies predict monotonically decreasing temperature profiles of the intracluster gas \citep{lok02}.  Observations of clusters with Chandra, such as Hydra A \citep{dm02}, however, indicate essentially flat temperature profiles.  The intracluster medium is dilute, magnetized, and has a mean free path that could be as high as one-tenth of the cluster virial radius.  It may be that these temperature profiles are representative of the saturated state of the MTI.  This possibility will be explored in future work.

\clearpage
\begin{figure}
\epsscale{.80}
\plotone{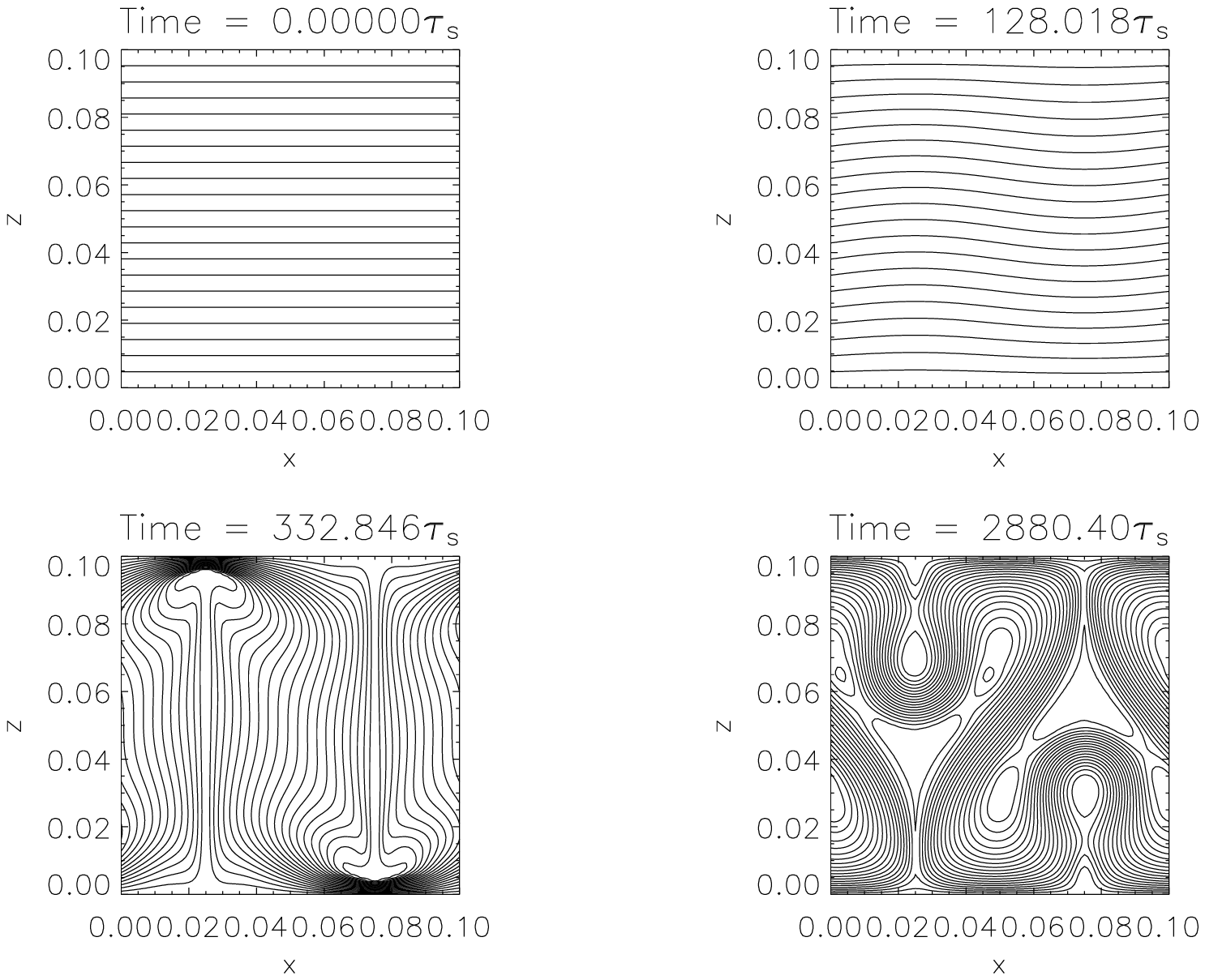}
\caption{Snapshots of the magnetic field lines for a single-mode perturbation with adiabatic boundary conditions  at various times
during the evolution of the instability. (upper left)
Inititial condition; (upper right) Linear phase; (lower left) Non-linear phase; (lower right) Saturated state.}

\end{figure}

\begin{figure}
\epsscale{0.80}
\plotone{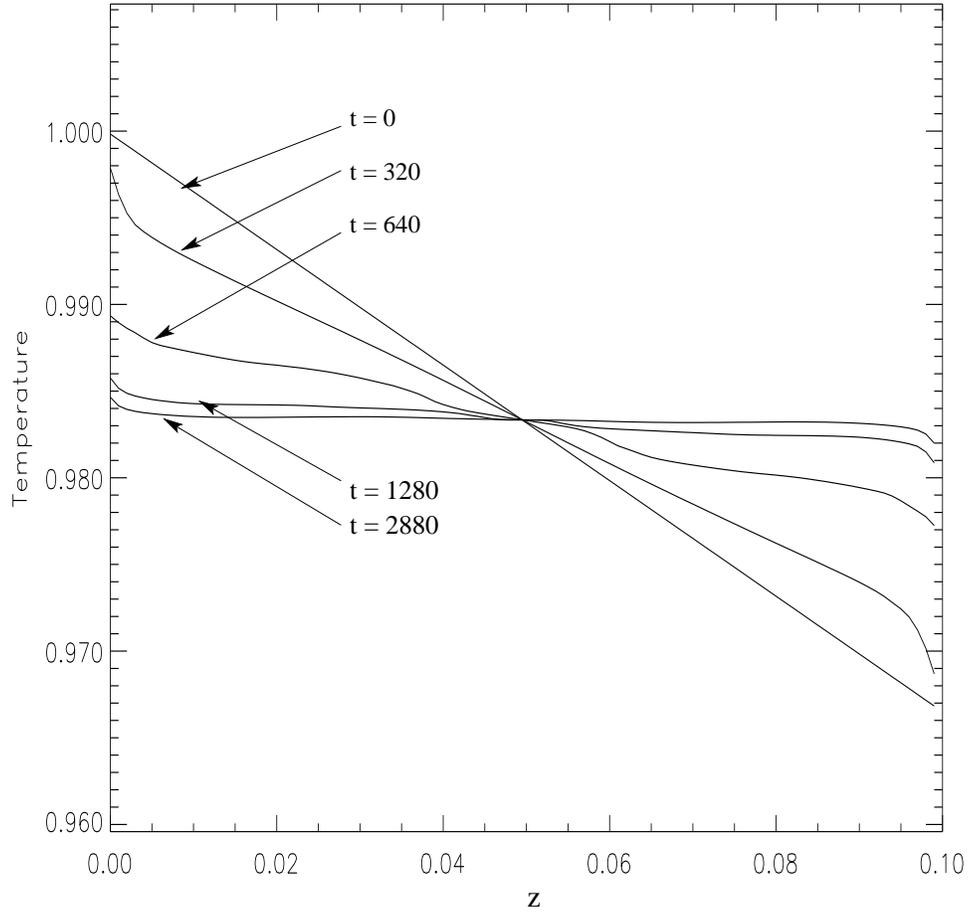}
\caption{Vertical profile of the horizontally-averaged temperature profile in the single mode case at various times.  The initial state
is a monotonically decreasing temperature profile with respect to height,
and the final state is isothermal.} \end{figure}

\begin{figure}
\epsscale{.80}
\plotone{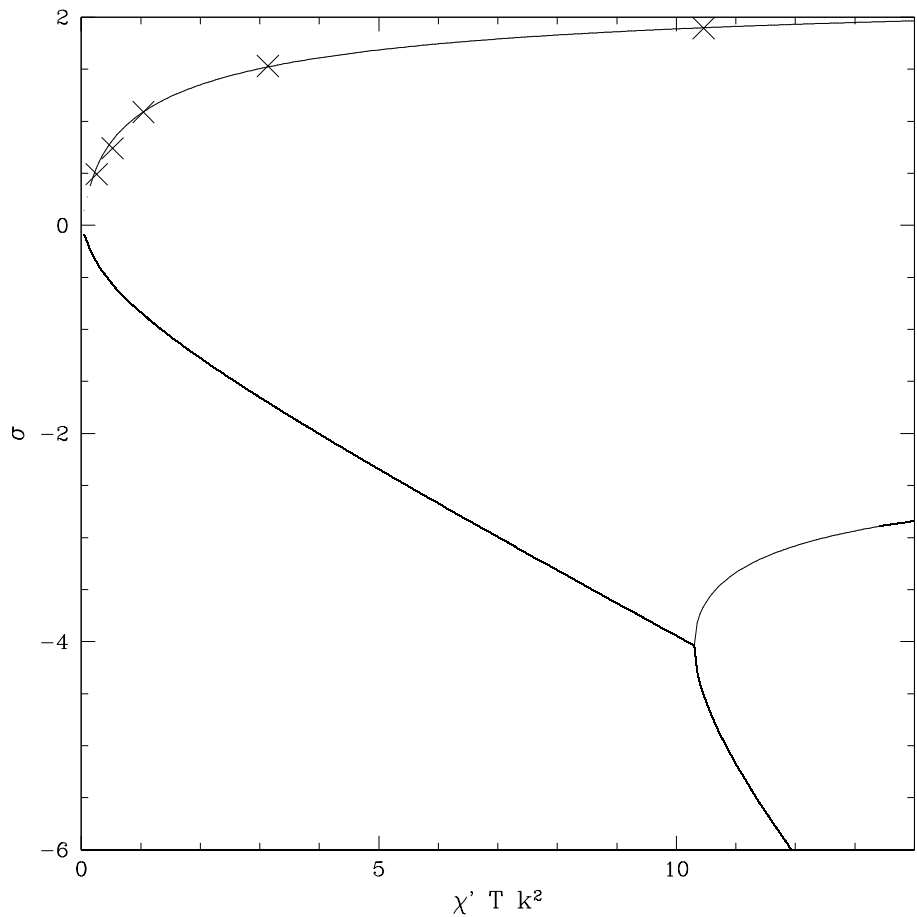}
\label{Plot_dispersion_relation}
\caption{Solutions of the MTI dispersion relation in the weak
field limit for an atmosphere with $d\ln T/d \ln S = -3.$  The axes
are normalized to the local Brunt-V\"{a}is\"{a}l\"{a} frequency, $N$.
The crosses are growth rates measured from simulations.}
\end{figure}

\begin{figure}
\epsscale{0.60}
\plotone{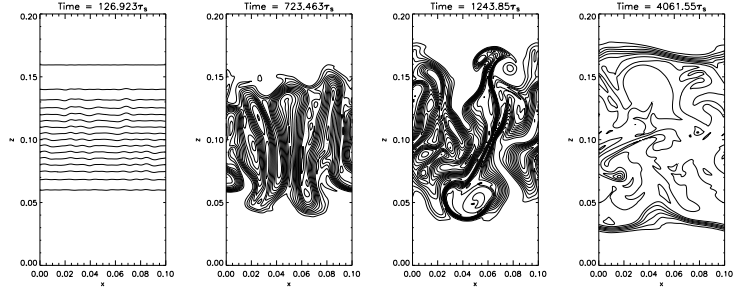}
\caption{Snapshots of the magnetic field in the run with stable layers.
(far left) Early linear phase;
(middle left) early non-linear phase. (middle right) The MTI drives penetrative convection into the
stable layers, and at late times (far right) magnetic flux is pumped into the stable
layers.} \end{figure}

\begin{figure}
\epsscale{0.60}
\plotone{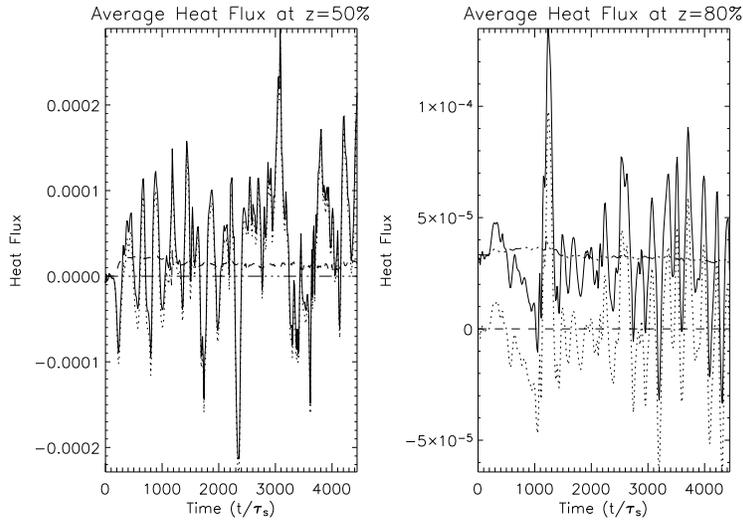}
\caption{Time evolution of the horizontally-averaged heat flux at the
midplane and 80\% height of the simulation domain in the run with stable layers.
The total heat flux
(thick solid line) is subdivided into Coulombic (thin solid line), radiative
(dashed lined), and advective (dotted line) components.  The instantaneous total heat flux is dominated by advective motions.} \end{figure}


\begin{thebibliography}{}
\bibitem[Balbus \& Hawley (1998)] {bh98} Balbus, S. A., \& Hawley J. F. 1998, Rev. Mod. Phys., 70, 1
\bibitem[Balbus (2000)] {bal00} Balbus, S. A. 2000, \apj, 534, 420
\bibitem[Balbus (2001)] {bal01} Balbus, S. A. 2001, \apj, 562, 909
\bibitem[Balbus (2004)] {bal04} Balbus, S. A. 2004, \apj, 616, 857
\bibitem[Braginskii (1965)] {brag65} Braginskii, S. I. 1965,  in Reviews of Plasma Physics, Vol. 1, ed. M. A. Leontovich (New York: Consultants Bureau), 205
\bibitem[Brummell, Clune, \& Toomre (2002)] {bru02} Brummell, N.H., Clune, \& T.L., Toomre, J. 2002, \apj, 570, 825
\bibitem[DeGrandi \& Modlendi (2002)] {dm02} DeGrandi, S., \& Molendi, S. 2002, \apj, 567, 163
\bibitem[Gardiner \& Stone (2005)]{gs05} Gardiner, T. \& Stone, J. 2005, J. Comp. Phys., 205, 509
\bibitem[Hammett \& Perkins (1990)]{hp90} Hammett, G. W. \& Perkins, F. W. 1990, \prl, 64, 3019
\bibitem[Loken, et al (2002)] {lok02} Loken, C., et al 2002, \apj, 579, 571
\bibitem[Markevitch (1998)] {mark98} Markevitch, M., et al. 1998, \apj, 377, 392
\bibitem[Parrish \& Stone (2005)] {ps05} Parrish, I. J., \& Stone, J. M., \apj 633, 334
\bibitem[Peterson \& Fabian (2005)] {pf05} Peterson, J. R., \& Fabian, A. C. 2005, astro-ph0512549
\bibitem[Quataert, Hammett, \& Dorland (2002)] {qdh02} Quataert, E., Dorland, W., \& Hammett, G. W. 2002 \apj, 577, 524
\bibitem[Sharma, Hammett, \& Quataert (2003)] {sha03} Sharma, P., Hammett, G., Quataert, E. 2003, \apj, 596, 1121
\bibitem[Spitzer (1962)]{spitz62} Spitzer, L. 1962, Physics of Fully Ionized Gases (New York: Wiley)
\bibitem[Tobias, et al. (2001)] {tob01} Tobias, S. M., et al. 2001, \apj, 549, 1183


\end{thebibliography}
\end{document}